\documentclass{llncs}

\usepackage[utf8]{inputenc}
\usepackage{amsmath}
\usepackage{amsfonts}
\usepackage[inference]{semantic} 
\usepackage{latexsym} 
\usepackage{stmaryrd} 
\usepackage{xcolor}
\usepackage{oz} 
\usepackage{wrapfig} 
\usepackage{breakcites} 
\usepackage{xspace}
\usepackage{tikz}
\usetikzlibrary{arrows,snakes,backgrounds,automata,shapes,positioning,chains,calc}
\usepackage{appendix}
\usepackage{hyperref}

\newcommand{\tid}{t}
\newcommand{\TS}{{\it TS}\xspace}

\newcommand{\rdval}{{\it rdval}}
\newcommand{\wrval}{{\it wrval}}

\newcommand{\var}{\mathit{var}}
\newcommand{\vmax}{\mathit{vmax}}

\newcommand{\interf}{interf}
\newcommand{\vps}{\beta}
\newcommand{\sync}{{\it sync}\xspace}
\newcommand{\fence}{fence}

\newcommand{\purple}[1]{{\color{purple}#1}}
\newcommand{\teal}[1]{{\color{teal}#1}}

\newcommand{\Wrt}{\mathit{Wr}}
\newcommand{\Rd}{\mathit{Rd}}

\newcommand{\wlp}{\operatorname{\mathsf{wlp}}}
\newcommand{\dis}{\operatorname{\mathsf{dis}}}

\newcommand{\Act}{{\sf Act}}
\newcommand{\Val}{{\sf Val}}

\newcommand{\Tid}{\mathsf{Tid}}
\newcommand{\VarG}{{\sf Var_G}}
\newcommand{\VarL}{{\sf Var_L}}
\newcommand{\Var}{{\sf Var}}

\definecolor{mycolor}{rgb}{0.122, 0.435, 0.698}

\newcommand{\linefill}{\cleaders\hbox{$\smash{\mkern-2mu\mathord-\mkern-2mu}$}\hfill\vphantom{\lower1pt\hbox{$\rightarrow$}}}  
  
\newcommand{\transi}[2]{\mathrel{\lower1pt\hbox{$\mathrel-_{\vphantom{#2}}\mkern-8mu\stackrel{#1}{\linefill_{\vphantom{#2}}}\mkern-11mu\rightarrow_{#2}$}}}

\newcommand{\tr}{traces}
\newcommand{\Acte}{{\sf Act_{ext}}}

\begin{document}

\title{View-Based Axiomatic Reasoning for PSO (Extended Version)\thanks{Bargmann and Wehrheim are supported by DFG-WE2290/14-1.}}

\author{Lara Bargmann \and
Heike Wehrheim}

\institute{Department of Computing Science, University of Oldenburg, Germany}

\maketitle
\begin{abstract}
Weak memory models describe the semantics of concurrent programs on modern multi-core architectures. 
Reasoning techniques for concurrent programs, like  Owicki-Gries-style proof calculi, have to be based on 
such a semantics, and hence need to be freshly developed for every new memory model. 
Recently, a more uniform approach to reasoning has been proposed which builds correctness proofs on the basis of 
a number of core {\em axioms}. This allows to prove program correctness {\em independent} of memory models, and 
transfers proofs to specific memory models by showing these to instantiate all axioms required in a proof. The axiomatisation is built 
on the notion of thread {\em views} as first class elements in the semantics.  

In this paper, we investigate the applicability of this form of axiomatic reasoning to the {\em Partial Store Order} (PSO) memory model. 
As the standard semantics for PSO is not based on views, we first of all provide a view-based semantics for PSO 
and prove it to coincide with the standard semantics. We then show the new view-based semantics to satisfy all but one axiom. 
The missing axiom refers to message-passing (MP) abilities of memory models, which PSO does not guarantee. As a consequence, only proofs without usage of the MP axiom are transferable to PSO. 
We illustrate the reasoning technique by proving correctness of a litmus test employing a fence to ensure message passing. 

\end{abstract}

\section{Introduction} \label{sec:intro} 
On multi-core architectures, the semantics of concurrent programs deviates from the often assumed {\em sequential consistency}   (Lamport~\cite{DBLP:journals/tc/Lamport79}). Sequential consistency guarantees that a concurrent program executes as an interleaving of statements following program order within threads. 
In contrast, the behaviour of concurrent programs on modern multi-core architectures looks like program statements have been reordered, e.g.~allowing for write-write reorderings (on disjoint variables). A {\em weak memory model} details the semantics of programs on such architectures. 

Today, weak memory models exist for a number of different architectures (e.g.~TSO~\cite{DBLP:conf/popl/SarkarSNORBMA09}, PSO and Power~\cite{DBLP:journals/computer/AdveG96}, ARM~\cite{DBLP:conf/popl/FlurGPSSMDS16}) as well as for programming languages (e.g.~C11~\cite{DBLP:conf/popl/BattyOSSW11}). The semantics is either specified in an {\em axiomatic} or an {\em operational} style. The operational style is more suitable for verification approaches like Hoare-style proof calculi~\cite{DBLP:journals/cacm/Hoare69} and their extensions to parallel programs by Owicki and Gries~\cite{DBLP:journals/acta/OwickiG76}  which need to construct proof outlines and reason about program statements. Recently, a number of operational semantics based on the concept of {\em views} have been proposed~\cite{DBLP:conf/ppopp/DohertyDWD19,ecoop20,DBLP:conf/ecoop/KaiserDDLV17,DBLP:conf/pldi/ChoLRK21,DBLP:conf/popl/KangHLVD17,DBLP:conf/esop/BilaDLRW22}. 
Views are specific to threads in concurrent programs and -- simply speaking -- specify the values of shared variables a thread can read (i.e., view) in some particular state\footnote{Note that in contrast to sequentially consistent execution, threads might be able to see different values of a shared variable in one state.}. 
View-based semantics lend itself well to Owicki-Gries style reasoning~\cite{DBLP:journals/acta/OwickiG76}, by replacing standard assertions on program variables by view-based assertions speaking about potential views of threads on shared variables. 
Still, with every new weak memory model, a new proof calculus for reasoning needs to be built. 

To alleviate this problem, Doherty et al.~\cite{TOCL} proposed a {\em generic} reasoning technique for weak memory models with a view-based operational semantics. The core of this technique are a number of axioms on the transition systems generated by the memory model. The axioms   detail properties of read and write actions (e.g., semi-commutativity of actions operating on different program variables) as well as fence instructions.  Correctness proofs of concurrent programs can then be done on the basis of axioms only,   independent of a concrete memory model. Proofs can be {\em transferred} to a specific memory model once the memory model has been shown to {\em instantiate} all axioms employed within the proof. 

In this paper, we investigate the applicability of this approach to the memory model {\em Partial Store Ordering} (PSO)~\cite{sparc}. A number of works have studied this memory model, proposing stateless model checking~\cite{DBLP:journals/acta/AbdullaAAJLS17}, proving the decidability of reachability~\cite{DBLP:conf/popl/AtigBBM10} or the NP-hardness of the testing (or consistency) problem~\cite{DBLP:journals/tecs/FurbachM0S15}. 
Besides an axiomatic semantics~\cite{DBLP:journals/fmsd/Alglave12}, PSO also has an operational semantics, however, not easily lending itself to the definition of views, in particular not for defining {\em view maximality}, the core concept underlying axiomatic reasoning in~\cite{TOCL}. Our first step is thus to develop a new semantics definition for PSO and prove it to coincide with the standard semantics via simulations~\cite{DBLP:journals/iandc/LynchV95}. Equipped with the new semantics, we prove PSO to instantiate all but one axiom of~\cite{TOCL}. The missing axiom refers to the ability of some memory models to provide {\em message passing} (MP) facilities, i.e.~to transfer the view of one thread $\tid_1$ to another thread $\tid_2$ when $\tid_2$ reads a value written by $\tid_1$. This property does not hold for PSO. Hence, only correctness proofs using the axioms other than MP are valid in PSO.
Besides the already existing proofs in~\cite{TOCL}, we provide a new correctness proof for a message passing example with a fence, required to guarantee correctness for weak memory models without message passing facilities, i.e.~requiring a proof without use of the MP axiom. On this, we exemplify the technique of axiomatic proving.

\begin{figure}[t] 
\footnotesize 
\begin{multicols}{2}
 $$\inference[\textsc{Exp}]{v=\llbracket e \rrbracket_{ls}}{(r:=e,ls)\overset{\tau}{\to} (skip,ls[r:=v])}$$
 $$\inference[\textsc{Write}]{a=wr(x,v) \text{ } v = \llbracket e \rrbracket_{ls}}{(x:=e,ls)\overset{a}{\to} (skip,ls)}$$ 
 $$\inference[\textsc{ParCom}]{(C_1,ls)\overset{a}{\to} (C_1',ls')}{(C_1;C_2,ls)\overset{a}{\to} (C_1';C_2,ls')} $$
 $$\inference[\textsc{If1}]{\llbracket b \rrbracket_{ls}}{(\text{if}\, b\, \text{then}\, C_1\, \text{else}\, C_2,ls)\overset{\tau}{\to} (C_1,ls)} $$
 
 $$\inference[\textsc{Read}]{a=rd(x,r,v) }{(r:=x,ls)\overset{a}{\to} (skip,ls[r:=v])}$$
 $$\inference[\textsc{Fence}]{a=fence}{(fnc,ls)\overset{a}{\to} (skip,ls)} $$
 $$\inference[\textsc{Skip}]{ }{(skip;C_2,ls)\overset{\tau}{\to} (C_2,ls)}$$
 $$\inference[\textsc{If2}]{\neg \llbracket b \rrbracket_{ls}}{(\text{if}\, b\, \text{then}\, C_1\, \text{else}\, C_2,ls)\overset{\tau}{\to} (C_2,ls)} $$
\end{multicols}

$$\inference[\textsc{While1}]{\llbracket b \rrbracket_{ls}}{(\text{while } b \text{ do } C,ls)\overset{\tau}{\to} (C;\text{while } b \text{ do } C,ls)} $$
$$\inference[\textsc{While2}]{\neg\llbracket b \rrbracket_{ls}}{(\text{while } b \text{ do } C,ls)\overset{\tau}{\to} (skip,ls)} $$
 
\vspace*{-0.3cm}    
 \caption{Local program semantics} 
 \vspace*{-0.3cm}
 \label{fig:locprosem} 
\end{figure}

\section{Background} \label{sec:background} 
Before looking at PSO and its view-based semantics, we define our program syntax and semantics partly following the notation of \cite{TOCL}.

\subsection{Program Syntax and Semantics} 
We define a concurrent program as a parallel composition of sequential programs. Each thread $t \in \Tid$ runs a sequential program $Com$ and with the function $\Pi:\Tid \to Com$ we model a concurrent program over threads $\Tid$. We let $\VarG$ be the set of global variables and $\VarL$ the set of local variables (or registers) with $\VarG \cap \VarL=\emptyset$ and $\Var=\VarG\cup \VarL$. We assume that initially all variables have the value 0. 

For $x\in \VarG$, $r\in \VarL$ and value $v\in \Val$ we define the following grammar
\begin{align*}
E 	&::= v \mid e\\
com &::= skip \mid fnc \mid r:=E \mid r:=x \mid x:=E \\
Com &::= com \mid Com;Com \mid \text{if } b \text{ then } Com \text{ else } Com \mid \text{while } b \text{ do } Com
\end{align*}
where $e\in Exp$ and $b\in BExp$ are expressions over local variables, $e$ arithmetic and $b$ boolean.

Alike a number of recent approaches for weak memory model semantics~\cite{TOCL,DBLP:conf/ppopp/DohertyDWD19,DBLP:conf/popl/LahavGV16},   
we define the semantics of a concurrent program running on a weak memory model by defining the semantics of programs {\em independent} of the memory model and later combining them with the memory model semantics. The program semantics first of all assumes that {\em any} value can be read for variables and the memory model later restricts these values. We will describe the operational rules for the memory model PSO (Fig.~\ref{fig:psosem}) below. 

First, we start with the set of actions 
$$\Act=\{rd(x,r,v),wr(x,v),\fence | x\in \VarG\land r\in \VarL \land v\in \Val \}$$
and add the read $r:=v$ and a silent action $\tau$ such that $\Acte=\Act \cup \{r:=v,\tau|r\in Var_L \land v\in \Val\}$. For an action $a\in\Act$ we will need the functions $\var,\rdval$ and $\wrval$, where $\var(a)\in \VarG$ describes the global variable of the action. If $a$ is a read action, then $\rdval(a)\in \Val$ describes its value, otherwise we set $\rdval(a)=\bot\notin \Val$. $\wrval(a)$ is defined in the same way for write actions. With these functions we can define the subsets of $\Act$: $Rd=\{a\in \Act | \wrval(a)=\bot \land \rdval(a)\neq \bot\}$ and $Wr=\{a\in \Act | \rdval(a)=\bot \land \wrval(a)\neq \bot\}$. For a value $v$, we assume $Rd[v]=\{a\in Rd | \rdval(a)=v\}$ and $Wr[v]=\{a\in Wr | \wrval(a)=v\}$. We let $\Act_{|x}$ be the set of all actions $a$ with $\var(a)=x$. Hence $Rd_{|x}$ is the set of all reads in $\Act_{|x}$ and $Wr_{|x}$ the set of all writes. 

The operational semantics for a sequential program is given by the transition rules in 
Figure~\ref{fig:locprosem}. Therein $ls : \VarL \rightarrow \Val$ defines the local state, which maps each local variable to its value, and $\llbracket e \rrbracket_{ls}$ is the value of the expression $e$ in the local state $ls$. By $ls[r:=v]$ we denote the local state which is equal to $ls$ for every local variable except for $r$ and the value of $r$ in $ls[r:=v]$ is   $v$. Note in particular that the {\sc Read} rule allows to use arbitrary values for $v$. To lift the sequential program of one thread to a concurrent program ($\Pi:\Tid\to Com$) we in addition employ the following rule 
\begin{align*}
    \inference[{\sc Par}]{(\Pi(t),lst(t))\overset{a}{\to} (C,ls) \text{ } a\in \Acte}{(\Pi,lst)\overset{a}{\to}_t (\Pi[t:=C],lst[t:=ls])}
\end{align*} 

\noindent where $lst: \Tid \to (\VarL \to \Val)$ maps each thread to its local state.

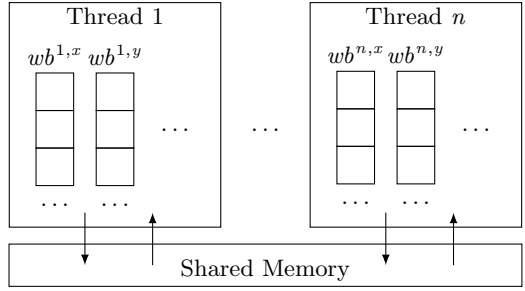
\begin{figure}[t]
\centering
    \footnotesize
    \begin{tikzpicture}
\tikzstyle{wb}=[draw,minimum size=0.5cm]

\begin{scope}[start chain=1 going below,node distance=-0.15mm]
    \node [on chain=1] at (-0.8,11.3) {$wb^{1,x}$};
    \node [on chain=1,wb] {};
    \node [on chain=1,wb] {};
    \node [on chain=1,wb] {};
    \node [on chain=1,wb,draw=none] {$\ldots$};    
\end{scope}

\begin{scope}[start chain=1 going below,node distance=-0.15mm] 
    \node [on chain=1] at (0,11.3) {$wb^{1,y}$} ;
    \node [on chain=1,wb] {};
    \node [on chain=1,wb] {};
    \node [on chain=1,wb] {};
    \node [on chain=1,wb,draw=none] {$\ldots$};    
\end{scope}

\begin{scope}[start chain=1 going below,node distance=-0.15mm] 
    \node [on chain=1,wb,draw=none] at (0.8,10.3) {$\ldots$};   
\end{scope}

\begin{scope}[start chain=1 going below,node distance=-0.15mm] 
    \node [on chain=1,wb,draw=none] at (2,10.3) {$\ldots$};   
\end{scope}        
       
\begin{scope}[start chain=1 going below,node distance=-0.15mm]
    \node [on chain=1] at (3.2,11.3) {$wb^{n,x}$};
    \node [on chain=1,wb] {};
    \node [on chain=1,wb] {};
    \node [on chain=1,wb] {};
    \node [on chain=1,wb,draw=none] {$\ldots$};    
\end{scope}

\begin{scope}[start chain=1 going below,node distance=-0.15mm] 
    \node [on chain=1] at (4,11.3) {$wb^{n,y}$} ;
    \node [on chain=1,wb] {};
    \node [on chain=1,wb] {};
    \node [on chain=1,wb] {};
    \node [on chain=1,wb,draw=none] {$\ldots$};    
\end{scope}

\begin{scope}[start chain=1 going below,node distance=-0.15mm] 
    \node [on chain=1,wb,draw=none] at (4.8,10.3) {$\ldots$};   
\end{scope}

\draw[] node[draw,minimum
        height=85,minimum width=80] (t1) at (0,10.5) {}; 
        \node[anchor=north,inner sep=2pt] at (t1.north)  {Thread 1};
        
\draw[] node[draw,minimum
        height=85,minimum width=80] (t2) at (4,10.5) {}; 
        \node[anchor=north,inner sep=2pt] at (t2.north)  {Thread $n$};

\draw[] node[draw,minimum
        height=16,minimum width=194] (s1) at (2,8.5) {}; 
        \node[anchor=south,inner sep=2pt] at (s1.south)  {Shared Memory};
        
\draw[-latex] (-0.4,9.2) 	-- (-0.4,8.5)	;
\draw[-latex] (0.5,8.5) 	-- (0.5,9.2)	;

\draw[-latex] (3.6,9.2) 	-- (3.6,8.5)	;
\draw[-latex] (4.5,8.5) 	-- (4.5,9.2)	;
\end{tikzpicture}
\vspace*{-0.3cm}
    \caption{PSO architecture}
    \vspace*{-0.5cm}
    \label{fig:pso}
\end{figure}

\subsection{PSO Semantics}
\begin{figure}[t] 
\centering
\footnotesize  
	$$\inference[\textsc{PSO-Read}]{ a=rd(x,r,v) \quad v=val_{\sigma}(t,x)   }
      {  \sigma \overset{a,t}{\leadsto}_{PSO} \sigma    }$$
	$$\inference[\textsc{PSO-Write}]{ a=wr(x,v) \quad wb'=\sigma .wb[(t,x) := \sigma .wb^{t,x}\cdot \langle v \rangle]   }
      {  \sigma \overset{a,t}{\leadsto}_{PSO} (s,wb')    }$$
	$$\inference[\textsc{PSO-Fence}]{ a=\fence \quad \sigma .wb^{t,x}= \langle \text{ } \rangle \text{ for all } x \in \VarG }
      {  \sigma \overset{a,t}{\leadsto}_{PSO} \sigma    }$$
    $$\inference[\textsc{PSO-Flush}]{ wb^{t,x} = \langle v \rangle \cdot w \quad
   	  s'= \sigma .s[x:=v] \quad wb'=\sigma .wb[(t,x):=w]   }
      {  \sigma \overset{flush,t}{\leadsto}_{PSO} (s',wb')    }$$
\vspace*{-0.5cm}
\caption{PSO semantics} 
\vspace*{-0.5cm}
 \label{fig:psosem} 
\end{figure}
Next, we give the operational semantics of PSO. This semantics is of an architectural style as it directly models architecture specific details. 
The PSO memory model contains a shared memory plus a write buffer per thread and per variable (see Fig.~\ref{fig:pso} for an illustration). 
The per-variable write buffers distinguish PSO from the TSO memory model~\cite{DBLP:journals/acta/AbdullaAAJLS17} which contains just one write buffer per thread. A write buffer $wb^{t,x}$ of a thread $t$ and a global variable $x$ is a FIFO ordered list with values as entries. Each write buffer can flush its first entry to the shared memory at any time. If a thread $t$ wants to read the value of a variable $x$, it either reads the last entry of $wb^{t,x}$ or, if $wb^{t,x}$ is empty, it reads from  shared memory. Next, we formally define the PSO semantics describing this behaviour.

In the PSO semantics, each state $\sigma=(s,wb)\in\Sigma_{PSO}$ contains a shared memory $s:\VarG\to \Val$ and a write buffer map $wb:(\Tid\times \VarG)\to \Val^*$, which maps each thread $t\in \Tid$ and each global variable $x\in \VarG$ to a write buffer list $wb^{t,x}$. With $I_{PSO}\subseteq \Sigma_{PSO}$ we denote the initial states, in which we assume $s(x)=0$ for all $x$ and write buffers to be empty. The rule \textsc{PSO-Write} in Figure \ref{fig:psosem} tells us that whenever a program writes some value $v\in \Val$ to a global variable $x$ in a thread $t$, we add $v$ to the write buffer $wb^{t,x}$. If we flush a write buffer $wb^{t,x}$ (rule \textsc{PSO-Flush}), we change the entry of $x$ in the shared memory $s$ to the first entry of $wb^{t,x}$ and delete it from the write buffer. When a program wants to read the value $v$ of a variable $x$ to a register $r$ in a thread $t$ (rule \textsc{PSO-Read}), it reads the last value of $wb^{t,x}$ in case the write buffer is not empty. Otherwise, the program reads the entry of $x$ directly from the shared memory. In the rule, this is described by the function $val_{\sigma}$. If the write buffer $wb^{t,x}$ is empty, then $val_{\sigma}(t,x) = \sigma.s(x)$, otherwise $val_{\sigma}(t,x) = last(\sigma .wb^{t,x})$. 
To pass a fence statement in a thread $t$ (i.e., to apply rule \textsc{PSO-Fence}) we first need to flush all write buffers of that thread until $wb^{t,x}$ is empty for every $x$.

To integrate the operational semantics of PSO in the program semantics, we will need to define the transition system $TS_{PSO}$ generated by programs running on PSO. Formally we write $TS_{PSO}=(\Act,\Sigma_{PSO},I_{PSO},T_{PSO})$ with the set of states $\Sigma_{PSO} = (\VarG \to \Val) \times ((\Tid \times \VarG) \to \Val^*)$, the set of initial states $I_{PSO} = \{ (s,wb)\in \Sigma_{PSO} |\forall t \in \Tid \text{ } \forall x \in \VarG : wb^{t,x} = \langle \text{ } \rangle \wedge  s(x)=0 \}$ and the set of transitions $T_{PSO}\in\Tid\times\Act\to 2^{\Sigma_{PSO}\times\Sigma_{PSO}}$. For an action $a\in\Act$ in a thread $t\in \Tid$, we set $T_{PSO}(t,a) = FL_{PSO}\fcmp \overset{a,t}{\leadsto}_{PSO}$ where $FL_{PSO}  = \left(\bigcup_{t\in \Tid}\overset{flush,t}{\leadsto}_{PSO}\right)^*$ and $\mathord{\fcmp}$ is the relational composition.

\subsection{Combined Semantics}
\label{sec:pso}
\begin{figure}[t]
  \centering
    \begin{minipage}[b]{0.45\columnwidth}
    \centering 
    \begin{tabular}{@{}l@{}}
    
  {\bf Init: } $x:=0;$ $y:=0;$ \\  
  $\begin{array}{@{}l@{\ }||@{\  }l@{}}
     \text{\bf Thread } 1
     & \text{\bf Thread } 2\\
     1: x := 1; \  &  3: r_1:=y; \\ 
     2: y:=  1; & 4: r_2:= x; \\
     \end{array}$
     \\
      {\color{green!40!black} $\big \{ r_1\in \{0,1\} \wedge r_2\in \{0,1\} \big \}$}  
   \end{tabular}
   \vspace*{-0.2cm}
 \caption{MP without fence}
 \vspace*{-0.5cm}
 \label{fig:mp}
 \end{minipage}
 \hfill
    \begin{minipage}[b]{0.45\columnwidth}
        \centering
    \begin{tabular}{@{}l@{}}

  {\bf Init: } $x:=0;$ $y:=0;$ \\ 
  $\begin{array}{@{}l@{\ }||@{\  }l@{}}
     \text{\bf Thread } 1
     & \text{\bf Thread } 2\\
     1: x:= 1; \ & 4: r_1:= y; \\
     2: fnc; \ & 5: r_2:=x; \\
     3: y:= 1; & 
     \end{array}$
\\
    \quad \quad   {\color{green!40!black} $\big \{ r_1=1 \implies r_2=1\big \}$}  
    \end{tabular}
    \vspace*{-0.2cm}
 \caption{MP with fence}
 \vspace*{-0.5cm}
 \label{fig:mp-fence}
 \end{minipage}
\end{figure}
Program and weak memory semantics are combined using the following three lifting rules.   
 $$ \inference[\textsc{Silent}]{(\Pi,lst)\overset{\tau}{\to}_t (\Pi',lst')}{(\Pi,lst,\sigma)\Rightarrow_t (\Pi',lst',\sigma)} 
 \quad
 \inference[\textsc{Local}]{(\Pi,lst)\overset{r:=v}{\longrightarrow}_t (\Pi',lst')}{(\Pi,lst,\sigma)\overset{r:=v}{\Longrightarrow}_t (\Pi',lst',\sigma)}$$

 $$\inference[\textsc{Memory}]{(\Pi,lst)\overset{a}{\to}_t (\Pi',lst') \quad a\in\Act \quad (\sigma, \sigma')\in T(t,a)}{(\Pi,lst,\sigma)\overset{a}{\Rightarrow}_t (\Pi',lst',\sigma')} $$
    
 The $\textsc{Memory}$-rule combines the rules for $a\in\Act$ of the local program semantics ($\textsc{Read}, \textsc{Write}, \textsc{Fence}$ and $\textsc{ParCom}$) with the ones of the PSO semantics. It is used for actions which affect the global memory. Here, $lst$ changes only if we read a value of a global variable to a register. We use the $\textsc{Local}$-rule to read a value directly to a register (\textsc{Read}), which changes $lst$, but since there is no global variable involved $\sigma$ stays the same. For the remaining rules of the local semantics we use the $\textsc{Silent}$-rule, which changes $lst$ only if we read an expression to a register ($\textsc{Exp}$).

We exemplify the semantics of PSO in the so-called {\em message passing} litmus test~\cite{litmus} in Figure~\ref{fig:mp}. It is called ``message passing'' as the expected behaviour is the following: when thread 2 reads $y$ to be 1, then it will afterwards read $x$ to be 1 as well. Thus, the message of $x$ to be 1 is passed from thread 1 to thread 2 upon reading $y$ to be 1. However, PSO also allows the outcome $r_1 = 1 \wedge r_2 = 0$ (no message passing). This can be explained as follows. 
First, thread 1 writes the value 1 to $x$, so $wb^{1,x}=\langle 1\rangle$. 
After that the same happens for $y$ and therefore $wb^{1,y}=\langle 1 \rangle$. 
Then just one write buffer is flushed before the reads, namely $wb^{1,y}$. In that case, thread 1 reads $y$ to be 1 from shared memory and $x$ to be 0. 
On the other hand, the outcome $r_1=1 \land r_2=0$ is not allowed in the message passing with fence example (Fig.~\ref{fig:mp-fence}). There we can only write the value of $y$ after the fence action, for which all write buffers of thread 1 have to be flushed. Hence thread 2 can only read the value 1 for $y$, if it also reads 1 for $x$. We will use the MP with fence example later to illustrate how our reasoning technique works. But first we explain axiomatic reasoning in the next section.

\section{Axiomatic Reasoning} \label{sec:axioms} 
\begin{figure}[t] 
   \begin{align*}
     \Sigma = {} &  \wlp(R,\Sigma) \tag{\text{Non-aborting}}
     \\
     R' \subseteq R \wedge P \subseteq P'  \imp {} &  \wlp(R,P) \subseteq \wlp(R',P')
                                            \tag{\text{(Anti)-Monotonicity}}
     \\
     \wlp(R,\wlp(R',P)) = {} &  \wlp(R \fcmp R',P) \tag{\text{Composition}}
     \\
     R[\wlp(R,P)] \subseteq {} &  P \tag{\text{Relation Application}}
     \\
     \wlp(R,P) \cap {} 
    \wlp(R,Q) = {} &   \wlp(R,P\cap Q) \tag{\text{Conjunctivity}}
     \\
     \wlp(R,P) \cup {} 
    \wlp(R,Q) \subseteq {} &   \wlp(R,P\cup Q) \tag{\text{Disjunctivity }}
   \end{align*}
   \vspace*{-0.8cm}
   \caption{Properties of wlp}
   \vspace*{-0.5cm}
   \label{fig:wlp} 
   \end{figure} 
The axioms of~\cite{TOCL} reason about arbitrary transition systems $TS \sdef (\Act,\Sigma,I,T)$ such as those of PSO, where $\Act$ is a set of actions, $\Sigma$ a set of states, $I \subseteq \Sigma$ a set of initial states and $T \in \Tid \times \Act \rightarrow 2^{\Sigma \times \Sigma}$ a set of transitions. The axiomatisation makes use of the {\em weakest liberal precondition transformer}~\cite{DBLP:books/ph/Dijkstra76}, as a basis for property specification and verification. 
For a relation $R$ and set of states $P$ (representing a predicate), let 
$\wlp : 2^{\Sigma \times \Sigma} \times 2^\Sigma \rightarrow 2^\Sigma$  
be 
$$\wlp(R,P) \sdef \{ \sigma \in \Sigma \mid \forall \sigma' : (\sigma,\sigma') \in R \implies \sigma' \in P \}
$$
 Some standard properties of $\wlp$ are given in Figure~\ref{fig:wlp}, where $\fcmp$ again denotes relational composition and $R[S]$ relational image. 
 
  In this work, $R$ is typically
instantiated to the relation $T(\tid, a)$, where $T$ is the transition relation, $\tid$   a thread and $a$   an action. We say $R$ is {\em disabled} in a state $\sigma$ iff $\sigma \in \dis(R)$ holds, where 
$\dis(R) \sdef \wlp(R, \emptyset)$. 

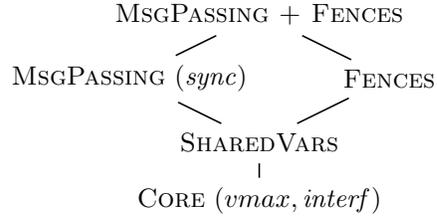
\begin{wrapfigure}{t}{0.5\textwidth} 
\vspace{-0.5cm}
\centering 
\begin{tikzpicture}[shorten >=1pt,auto,node distance=1.2cm,
                    semithick]
  \tikzstyle{every state}=[scale=.65,initial text=]

  \node (A) { {\sc Core} ($\vmax, \interf$)}; 
  \node [above of=A,yshift=-12pt] (B) { {\sc SharedVars}} ; 
  \node [above left of=B,xshift=-25pt] (C) { {\sc MsgPassing} ($sync$)};
  \node [above right of=B,xshift=25pt] (D) {{\sc Fences}}; 
  \node [above right of=C,xshift=25pt] (E) {{\sc MsgPassing + Fences}};
  \path (A) edge (B)
        (B) edge (C)
        (B) edge (D)
        (C) edge (E)
        (D) edge (E); 
  \end{tikzpicture}
\vspace*{-0.4cm}
\caption{Axiom Hierarchy}
\vspace*{-0.7cm}
\label{fig:hierarchy}
\end{wrapfigure} 

Next, we give the axioms of~\cite{TOCL}. The axioms are structured in a hierarchy in which every level adds a set of axioms to the axioms of lower levels.  Figure~\ref{fig:hierarchy} gives the hierarchy. 
When instantiating a memory model, parameters $\vmax, \interf$ (of level {\sc Core}) and $\sync$ (of level {\sc MsgPassing}) have to be concretised. Their meaning is described below. 

\paragraph{Level {\sc Core}.} 
The lowest level contains basic axioms describing core properties of view-based memory model semantics. The key concept here is that of {\em view maximality}: a thread is view maximal (on an action operating on a variable $x\in \VarG$) if it has the ``most up-to-date" view on this variable. In PSO, a thread $\tid$ is view maximal on e.g.~reading $x$, if either $x$ is in shared memory and in no write buffer or $\tid$'s write buffer for $x$ has an entry which will be flushed to shared memory ``later'' than all other values for $x$. 
This means, while non view maximal threads might be able to read older values of $x$, thread $\tid$ reads the most up-to-date value. 
However, as this concept of ``being flushed later'' is not present in the semantics of PSO we will later need to define an alternative semantics. This will enable us to instantiate $\vmax$ for PSO.

The core ingredient of the axiom system are view-preserving simulations. 

\begin{definition}\label{def:beta}
For a transition system $\TS=(\Act,\Sigma,I,T)$, a {\em view-preserving simulation},
denoted $\vps$, is the weakest relation $R$ satisfying the following, for all threads
$\tid \in \Tid$ and all actions $a \in \Act$ 
\begin{align}
  \tag{\text{semi-commutation}}
  \label{eqn:vps1}
R \fcmp T (t, a) & \subseteq T (t, a) \fcmp R\\
  \tag{\text{view maximality}}
\label{eqn:vps2}
\vmax(t, a) & \subseteq \wlp(R, \vmax(t, a))
\end{align}
\end{definition}

View-preserving simulations conceptionally include all sorts of system steps changing the state without executing visible actions like read and write. In PSO, these would be flushes. In order for these steps to not mess up the proofs, the relation associated with these steps should keep view maximality of threads and should semi-commute with the transition relation. 

Furthermore, the core axioms make use of an {\em interference relation} $\interf \in \Tid \times \Act \to 2^{\Sigma  \times \Sigma}$ which (together with $\vps$) provides an overapproximation of the transition relation $T(\tid,a)$ in order to abstract from details of the memory model and to regain standard properties of reasoning (like writes and reads on different variables commuting). 
Figure~\ref{fig:core-axioms} gives all core axioms.

For reasoning we will later employ the following {\em view-based} assertions.  

\begin{definition} 
\label{def:assertions}
For a thread $t$, a variable $x \in \VarG$ and a value $v \in \Val$ we define

\vspace{.5em}
$\begin{array}{@{}r@{~}l@{\ \ }l} 
    [x \not \approx v]_t &\sdef \dis(T(t, \Rd_{|x}[v])) & \text{(Impossible value)} \\  
    {[}x \equiv v]_t 
                         &\sdef \bigcap_{u \neq v} [x \not \approx u]_t  &\text{(Definite value)}\\
    x_{\uparrow t}  & \sdef \bigcap_{a\in \Act_{|x}} \vmax(t,a) 
    & \text{(Maximal view)} \\ 
   {[}x = v]_t
&\sdef {[}x \equiv v]_t \cap x_{\uparrow t} & \text{(Synced value)} 
\end{array}$
\end{definition}

\noindent Assertions should be (and indeed are) stable under $\vps$. 

\begin{definition}
Predicate $P \in 2^\Sigma$ is {\em $\vps$-stable} iff
$P \subseteq \wlp(\vps,P)$. 
\end{definition}

In fact, the four assertions introduced above are $\vps$-stable. Furthermore, the following property is derivable from the axioms.  

\begin{lemma}[\cite{TOCL}]\label{lem:overapprox}  
    For any $a \in \Act$, thread $t$ and $\vps$-stable predicate $P$, if 
    $P \subseteq \wlp(\interf(t,a),P)$, then $P \subseteq \wlp(T(t,a),P).$
\end{lemma}

\begin{figure}[t]
 \begin{align*}    
     \textbf{C1}: & \forall \tid \in \Tid, a \in \Act: I \subseteq \vmax(\tid, a)  \\
     \textbf{C2}: & \forall \sigma, \sigma' \in \Sigma, \tid, \tid' \in \Tid, a \in \Act: \\ 
    & \sigma \in \vmax(\tid,a) \wedge (\sigma,\sigma') \in T(\tid,a) \imp  \exists
  \tau \in \Sigma:\ (\sigma',\tau) \in \vps \wedge (\sigma,\tau) \in T(\tid',a) \fcmp\vps \\
    \textbf{C3}: & \forall \tid \in \Tid, a\in \Act:  
     T (\tid, a) \subseteq \vps \fcmp \interf (\tid, a) \fcmp \vps \\
  \textbf{C4}: & \forall \tid \in \Tid, a,b \in \Act:  
      \vmax (t, a) \subseteq \wlp(\interf (t, b),\vmax (t, a))
   \end{align*}
   \vspace*{-0.8cm}
    \caption{The axioms of  level {\sc Core}}
    \vspace*{-0.5cm}
    \label{fig:core-axioms}
\end{figure}
\paragraph{Level {\sc SharedVars}.} 
\begin{figure}[t]
	\begin{align*}    
		\textbf{SV1}: &  \forall a,b \in \Act, \tid, \tid'\in \Tid \text{ s.t. } \var(a) \neq \var(b): \\ 
		& \interf (\tid', b) \fcmp T (\tid, a) \subseteq T (\tid, a) \fcmp \interf (\tid', b) \\
		\textbf{SV2}: & \forall a,b \in \Act, t,t'\in \Tid \text{ s.t. } \var(a) \neq \var(b):  \\
		& \vmax (t,a) \subseteq \wlp(\interf (t', b), \vmax (t,a)) \\
		\textbf{RW1}: & \forall t,t' \!\in\! \Tid, x \!\in\! \VarG, a_r\! \in\! \Rd_{|x}, a_w\! \in\! \Wrt_{{|}x} \text{ s.t. } \rdval(a_r) \not =  \wrval(a_w): \\ 
		& \interf(t', a_w) \fcmp T(t, a_r) \subseteq T(t, a_r) \fcmp \interf(t', a_w) \\
		\textbf{RW2}: & \forall a\in \Act, t,t'\in \Tid, a_r \in \Rd_{|\var(a)}:\\ & \interf(t', a_r) \fcmp T(t,a) \subseteq T(t,a) \fcmp \interf(t', a_r) \\
		\textbf{RW3}: & \forall a \in \Act, t,t'\in \Tid, a_r \in \Rd_{|\var(a)}:\\  & \vmax (t, a) \subseteq \wlp(\interf(t', a_r), \vmax (t, a)) \\
		\textbf{RW4}: & \forall x\in \VarG, t \in \Tid:  \Sigma \subseteq \dom (T (t, Rd_{|x})) \\
		\textbf{RW5}: & \forall x \in \VarG, a_w \in \Wrt_{|x}, v = \wrval(a_w):  \Sigma \subseteq \wlp(T(t, a_w), \dom (T (t, \Rd_{|x}[v]))) \\
		\textbf{RW6}: & \forall x\in \VarG, t \in \Tid, v\in \Val: 
		x_{\uparrow t} \subseteq \bigcup_{v \in \Val} {[}x \equiv v]_t  \\
		\textbf{RW7}: & \forall x \in \VarG, a_w, a_r, a\in \Act_{|x}, t, t' \in \Tid, \text{s.t.}\, \wrval(a_w) = \rdval(a_r) \wedge t \neq t':\\&
		\vmax(t, a_w) \cap \dis(T(t',a_r)) \subseteq \wlp(T(t,a_w), \wlp(T(t', a_r), \vmax(t', a)))
	\end{align*} 
	\vspace*{-0.8cm}
	\caption{The axioms of  level {\sc SharedVars}} 
	\vspace*{-0.5cm}
	\label{fig:sv-axioms}
\end{figure}

The next level of the axiom hierarchy deals with axioms about actions with respect to the shared variables they access. 
The level \textsc{Shared Variables} contains all axioms of the level \textsc{Core} plus those given in Figure~\ref{fig:sv-axioms}. We exemplarily look at two of them, namely \textbf{SV1} and \textbf{RW6}. 
The axiom \textbf{SV1} states a semi-commutation property of actions (i.e., reads and writes) on {\em different} variables. This is a property commonly expected for programming languages, e.g.~if we first write 4 to $x$ and then 5 to $y$ and $y \neq x$, then we should reach the same state if we first write 5 to $y$ and then 4 to $x$. However, this does not necessarily hold for weak memory models. For instance, in TSO~\cite{DBLP:conf/popl/SarkarSNORBMA09}  with just one write buffer for all locations we would reach two different states. To regain this property for axiomatic reasoning, the axiom uses $\interf(t',b)$ (instead of $T(t',b)$).

Axiom \textbf{RW6} uses the notion of {\em maximal view} of a thread $\tid$ on a variable $x \in \VarG$. 
\textbf{RW6} states that a thread being maximal wrt.~$x$ has to know a definite value for $x$.  

When the axioms \textbf{C3}, \textbf{SV1} and \textbf{SV2} hold, we get the following property for $\beta$-stable predicates:

\begin{lemma}\label{lem:fencestable} \ \\
    For all $P\subseteq\Sigma$ and threads $t$, if $P\subseteq \wlp(\beta, P)$, then $P\subseteq \wlp(T(t,\fence),P)$.
\end{lemma}

If additionally \textbf{RW2} and \textbf{RW3} hold, we can show the same for read actions.

\begin{lemma}[\cite{TOCL}]\label{lem:readstable}
    For all $P\subseteq\Sigma$, threads $t$ and $a_r\in \Rd$, if $P\subseteq \wlp(\beta, P)$, then $P\subseteq \wlp(T(t,a_r),P)$.
\end{lemma}

\paragraph{Levels {\sc MsgPassing} and {\sc Fences}.} 
Finally, we have two levels with just one additional axiom (plus a level {\sc MsgPassing + Fences} uniting these two). Level {\sc MsgPassing} contains all axioms of {\sc SharedVars}  plus the following message passing axiom. 

\begin{description}
   \item[MP] 
  For $a_w,a_r, b \in \Act$ and $\purple{t},\teal{t'} \in \Tid$ such that $(a_w,a_r) \in \sync$, $\var(a_w) = \var(a_r)$,  $\wrval(a_w) = \rdval(a_r)$, $\var(b) \neq \var(a_w)$, and $\purple{t} \neq \teal{t'}$, we have

  \begin{align*}
      &\vmax(t, b) \cap \wlp(\teal{T(t', a_r)}, \vmax (t', b))\\ 
      &\subseteq
      \wlp(\purple{T(t, a_w)}, \wlp(\teal{T(t', a_r)}, \vmax (t', b))). 
  \end{align*}

\end{description} 

Intuitively, \textbf{MP} states the following property: we consider two actions $a_w$ and $a_r$, the first a write and the second a read. The read action reads the value written by $a_w$. Furthermore, $(a_w,a_r) \in \sync$ means that reading provides synchronisation with writing. The relation $\sync$ is a parameter to the axiomatisation which -- like $\interf$ and $\vmax$ -- needs to be instantiated for a concrete memory model. As some examples: in TSO, any write is in $\sync$ with any read; in C11 RAR~\cite{DBLP:conf/ppopp/DohertyDWD19} we only have writes marked as \textsc{Releasing} in sync with reads marked as \textsc{Acquiring}. 

The axiom then requires the following: if thread $t$ is view maximal on action $b$, say on variable $x$, and the read action $a_r$ would make thread $t'$ view maximal on $b$ as well, then after performing the write $a_w$ and the read $a_r$ thread $t'$ actually becomes view maximal on $b$\footnote{This is akin to the Shared-Memory Causality Principle of Lahav and Boker~\cite{DBLP:journals/toplas/LahavB22}.}. 

Note that this axiom does not hold for PSO (or, rather we could make it hold by setting $\sync$ to $\emptyset$ and then not being able to apply it anywhere). The reason for this is the use of separate write buffers per variable as already explained on the example in Figure~\ref{fig:mp}. 

The last axiom of level {\sc Fences} deals with fence instructions. It states that a thread $t$ being view maximal on $a$ makes all other threads view maximal on $a$ when executing a fence instruction. 

\begin{description}
   \item[FNC]
     $\forall a\in \Act,\, t,t' \in \Tid$: $\vmax(t,a) \subseteq \wlp(T(t,\fence),\vmax(t',a))$.
\end{description} 

\noindent 
Finally, to reason about assignments to local registers $r \in \VarL$ a standard wlp rule is required:
\begin{eqnarray}
    \label{eq:subs-rd}
  \mathit{e}[r := v] & \subseteq & \wlp(T(\tid,rd(x,r,v)),\mathit{e}) 
 \end{eqnarray}
Here, $\mathit{expr}[r := v]$ is the replacement of $r$ by $v$ within expression $e$. 
This completes the axiom set. Next, we look at PSO again, instantiate the parameters $\vmax$ and $\interf$ for PSO and then prove PSO to satisfy all axioms of level {\sc SharedVars} plus axiom \textbf{FNC}. 
\section{Prophetic PSO}
\label{sec:ppso}
\begin{figure}[t] 
\footnotesize 
	$$\inference[\textsc{PP-Read}]{ a=rd(x,r,v) \quad v=val_{\sigma}(t,x)   }
      {  \sigma \overset{a,t}{\leadsto}_{PP} \sigma    }$$ 
	$$\inference[\textsc{PP-Write}]{ a=wr(x,v) \quad fresh_{\sigma}(t,x,q)\quad
   wb'=\sigma .wb[(t,x) := \sigma .wb^{t,x}\cdot \langle (v,q) \rangle]  }
      {  \sigma \overset{a,t}{\leadsto}_{PP} (s,wb')    }$$
	$$\inference[\textsc{PP-Fence}]{ a=\fence \quad \sigma .wb^{t,x}= \langle \text{ } \rangle \text{ for all } x \in \VarG   }
      {  \sigma \overset{a,t}{\leadsto}_{PP} \sigma    }$$
	$$\inference[\textsc{PP-Flush}]{ wb^{t,x} = \langle (v,\cdot) \rangle \cdot w \quad s'= \sigma .s[x:=v]\\
   	  nextFlush_{\sigma}(t,x) \quad wb'=\sigma .wb[(t,x):=w]   }
      {  \sigma \overset{flush,t}{\leadsto}_{PP} (s',wb')    }$$
\vspace*{-0.5cm}
\caption{PPSO semantics}
\vspace*{-0.5cm}
 \label{fig:ppsem} 
\end{figure}

To use our axiomatisation, we need to instantiate the parameters $vmax$ and $interf$ for the concrete memory model. However, this is not straightforward for PSO as we cannot see view maximality from the state; we do not know for which thread the entries of the write buffers are being flushed last, so we do not know which thread has the ``most up-to-date" view on a variable. To still be able to use the axioms for reasoning on PSO, we thus first provide an alternative semantics for PSO called {\em Prophetic PSO} (PPSO). It is ``prophetic" in the sense of knowing the order of flushing entries of write buffers. To this end, we directly assign a timestamp to every entry when writing and only flush entries in order of timestamps. PPSO is then shown to be trace equivalent to PSO.  

The semantics of PPSO is given in Figure~\ref{fig:ppsem}. PSO and PPSO only differ in the type of write buffers. In the prophetic version, they do not only contain the values written, but also a timestamp $q\in \mathbb{Q}$ for each write. Therefore, we can only write a value $v$ to a variable $x$ in a thread $t$, if the condition $fresh_{\sigma}(t,x,q)=true$ in the current state $\sigma$. Formally $fresh_{\sigma}(t,x,q)$ is defined as 
$$(\forall t' \in \Tid \text{ } \forall x' \in \VarG : (\cdot,q) \notin \sigma .wb^{t',x'}) \wedge (\forall (\cdot,q') \in \sigma .wb^{x,t} : q>q')$$
This means that none of the write buffers contains the timestamp $q$ and all timestamps in $wb^{t,x}$ are smaller than $q$. In that case, we can add the pair $(v,q)$ to $wb^{t,x}$. To flush the first entry $(v,q)$ of a write buffer $wb^{t,x}$ the condition $nextFlush_{\sigma}(t,x)$ has to be $true$. That is the case if 
$$(\exists q : (\cdot,q) = wb^{t,x}(0)) \wedge (\forall t' \forall x' \forall (\cdot,q') \in wb^{t',x'} : t' \neq t \vee x' \neq x \Rightarrow q'>q)$$
Then $q$ is smaller then every timestamp from every other write buffer. Then we can delete the entry and change the memory as before. The read and fence rules are the same as the ones in the PSO semantics. Only $val_{\sigma}(t,x)$ is now the value of the last entry of $\sigma.wb^{t,x}$, if $wb^{t,x}$ is not empty. In this case we write $val_{\sigma}(t,x)=val(last(\sigma .wb^{t,x}))$, otherwise $val_{\sigma}(t,x)$ is still $\sigma.s(x)$.  

We define the transition system $T_{PP}$ that is generated by programs running on PPSO analogous to $TS_{PSO}$. Hence we set $TS_{PP}=(\Act,\Sigma_{PP},I_{PP},T_{PP})$ with the set of states $\Sigma_{PP} = (\VarG \to \Val) \times ((\Tid \times \VarG) \to (\Val \times \mathbb{Q})^*)$, the set of initial states $I_{PP}	= \{ (s,wb) | s \in \VarG \to \Val \wedge \forall t \in \Tid \text{ } \forall x \in \VarG : wb^{t,x} = \langle \text{ } \rangle \}$ and the set of transitions $T_{PP}\in \Tid\times\Act\to 2^{\Sigma_{PP}\times\Sigma_{PP}}$. Similarly to $T_{PSO}$, we set $T_{PP}(t,a)=FL_{PP}\fcmp\overset{a,t}{\leadsto}_{PP}$ where $FL_{PP}=\left(\bigcup_{t\in \Tid}\overset{flush,t}{\leadsto}_{PP}\right)^*$.

We call this semantics {\em prophetic PSO} because with the presence of timestamps we can now predict the  order of write buffer flushes. Therefore, we know whether a state $\sigma\in \Sigma_{PP}$ has the maximal view for an action $a$ in a thread $t$. With that we can define $\vmax$ and $\interf$ for PPSO and investigate which axioms from Section~\ref{sec:axioms} hold. First we however prove that PSO and PPSO (or more precisely $TS_{PSO}$ and $TS_{PP}$) are {\em trace equivalent}. Every correctness proof for a program running on PPSO then also holds for PSO.  

For showing trace equivalence, we first need some definitions. Here we partly follow the notation of Lynch and Vaandrager~\cite{DBLP:journals/iandc/LynchV95}. 
Let $TS_A=(\Act,\Sigma_A,I_A,T_A)$ and $TS_C=(\Act,\Sigma_C,I_C,T_C)$ be two transition systems. A trace of $TS_A$ is a finite sequence of actions
$\alpha\in \Act^*$ with $\exists\sigma_0\in I_A,\sigma\in\Sigma_A: \sigma_0 \overset{\alpha}{\Rightarrow} \sigma$. Here $\sigma_0 \overset{\alpha}{\Rightarrow} \sigma$ means $(\sigma_0,\sigma)\in T_A(t_1,a_1)\fcmp...\fcmp T_A(t_n,a_n)$ for $\alpha=a_1...a_n\in\Act^*$ and $t_1,...,t_n\in\Tid$. By $\tr(TS_A)$ we note all traces of $TS_A$. We call two transition systems $TS_A$ and $TS_C$ {\em trace equivalent} if $\tr(TS_A) = \tr(TS_C)$. Hence, to show that $TS_{PSO}$ and $TS_{PP}$ are trace equivalent, we prove that (i) $\tr(TS_{PP}) \subseteq \tr(TS_{PSO})$ and (ii) $\tr(TS_{PSO}) \subseteq \tr(TS_{PP})$. 
To do so, we employ forward and backward simulations. 
We start with forward simulations and the proof of (i).

\begin{definition}[\cite{DBLP:journals/iandc/LynchV95}]
\label{def:fsim}
A forward simulation from $TS_A$ to $TS_C$ 
is a relation $F\subseteq \Sigma_A \times \Sigma_C$ with $F(\sigma_0)\cap I_C\neq \emptyset$ for all $\sigma_0\in I_A$ such that for every $a\in\Act$

\begin{align*}
    (\sigma_1,\sigma_2)\in F \land (\sigma_1, \sigma'_1)\in T_A(t,a) \Rightarrow \exists \sigma'_2 : (\sigma_2,\sigma'_2)\in T_C(t,a) \wedge (\sigma'_1,\sigma'_2)\in F.
\end{align*}
\end{definition}
Lynch and Vaandrager \cite{DBLP:journals/iandc/LynchV95} tell us that whenever there exists a forward simulation from $TS_{PP}$ to $TS_{PSO}$, then $\tr(TS_{PP})\subseteq \tr(TS_{PSO})$. Therefore, we need the following theorem:
\begin{theorem}\label{th:fsim}
There exists a forward simulation from $TS_{PP}$ to $TS_{PSO}$. 
\end{theorem}

In the proof of this theorem, we choose a relation $F\subseteq\Sigma_{PP}\times\Sigma_{PSO}$ and show that $F$ is a forward simulation. For $\sigma_1=(s_1,wb_1)$ and $\sigma_2=(s_2,wb_2)$ we say 
\begin{align*}(\sigma_1,\sigma_2)\in F : \Leftrightarrow s_1=s_2 \wedge \forall t\in \Tid \text{ } \forall x\in \VarG  : wb_2^{t,x}= wb_1^{t,x}.val\end{align*}
With $wb^{t,x}.val$ we describe a write buffer which contains only the values of $wb^{t,x}$. This means our relation contains only pairs that have the same shared memory and the same values in every write buffer. The full proof can be found in the appendix. (i) follows directly from this theorem. Next we look at backward simulations to prove (ii).
\begin{definition}[\cite{DBLP:journals/iandc/LynchV95}]
\label{def:bsim}
A backward simulation from $TS_A$ to $TS_C$ is a relation $B\subseteq \Sigma_A \times \Sigma_C$ with $B(\sigma_0)\subseteq I_C$ for all $\sigma_0\in I_A$ such that for every $a\in \Act$
\begin{align*}
    (\sigma'_1,\sigma'_2)\in B \land  (\sigma_1, \sigma'_1)\in T_A(t,a)  \Rightarrow \exists \sigma_2 : (\sigma_2,\sigma'_2)\in T_C(t,a) \wedge (\sigma_1,\sigma_2)\in B
\end{align*}
and $B$ is total on $\Sigma_A$.
\end{definition}

In the next theorem we show that there exists a backward simulation from $TS_{PSO}$ to $TS_{PP}$.
Since we only look at finite traces, \cite{DBLP:journals/iandc/LynchV95} proved that then $\tr(TS_{PSO})\subseteq \tr(TS_{PP})$.

\begin{theorem}\label{th:bsim}
There exists a backward simulation from $TS_{PSO}$ to $TS_{PP}$. 
\end{theorem}

We prove this theorem by choosing the relation $B\subseteq \Sigma_{PSO}\times\Sigma_{PP}$ analogous to the forward simulation with 
\begin{align*}
(\sigma_1,\sigma_2)\in B : \Leftrightarrow s_1=s_2 \wedge \forall t\in \Tid \text{ } \forall x\in \VarG  : wb_1^{t,x}= wb_2^{t,x}.val
\end{align*}
for $\sigma_1=(s_1,wb_1)\in\Sigma_{PSO}$ and $\sigma_2=(s_2,wb_2)\in\Sigma_{PP}$.
To show that $B$ is a backward simulation, we use the following invariant:

\begin{lemma}\label{lem:bsim}
Let $t\in \Tid$, $x\in \VarG$, $\sigma_0\in I_{PP}$ and $\sigma\in \Sigma_{PP}$ with $\sigma_0\overset{\alpha}{\Rightarrow} \sigma$ for $\alpha\in \tr(TS_{PP})$. If $q=ts(last(\sigma.wb^{t,x}))$ we have
\begin{align*}
\forall t' \in \Tid \text{ } \forall x' \in \VarG,t'&\neq t \lor x'\neq x: (\cdot,q) \notin \sigma .wb^{t',x'}\\ \land\forall (\cdot,q') \in \sigma .wb^{x,t},(\cdot,&q')\neq last(\sigma.wb^{t,x}): q>q'
\end{align*}
\end{lemma}
We will need that lemma to ensure that $fresh_{\sigma}(t,x,q)=true$ for the last timestamp $q$ of every write buffer $wb^{t,x}$ in every reachable state $\sigma\in\Sigma_{PP}$. Hence, we can execute a write action under PPSO for every variable in every thread at any time. Both proofs can be found in the appendix. 
\begin{corollary}\label{cor:treq}
$TS_{PSO}$ and $TS_{PP}$ are trace equivalent.
\end{corollary}

\begin{figure}[t] 
\footnotesize 
\begin{align*}
\vmax_{PP}(t,a)=	& \begin{cases} maxTS(t,var(a)), & \text{if } a\in Wr\cup Rd \\
					\Sigma_{PP}, 					 & \text{else} \end{cases}\\
\interf_{PP}(t,a)=	& \begin{cases} id,         &\text{if } a\in Rd\cup \{\fence\} \\
					T_{PP}(t,a)\fcmp FL_{PP},	& a\in Wr \end{cases} \\
maxTS(t,x)=			&\{\sigma | \sigma.wb^{t,x}\neq \langle \text{ } \rangle \wedge (\forall t',q:(\cdot,q)\in\sigma.wb^{t',x}\Rightarrow ts(last(\sigma.wb^{t,x}))\geq q)\}\\
					& {} \cup\{\sigma | \forall t':\sigma.wb^{t',x}= \langle \text{ } \rangle\}
\end{align*}
\vspace*{-0.8cm}
\caption{PPSO definitions }
\vspace*{-0.5cm}
 \label{fig:ppdef} 
\end{figure}
With that equivalence we can show on which level PSO fits in the axiom hierarchy, by showing which axioms hold for PPSO. To do so, we first need to instantiate the parameters mentioned in Section~\ref{sec:axioms}. 
Figure~\ref{fig:ppdef} gives an overview of these definitions. 
The set of states with maximal view $\vmax_{PP}(t,a)$ for instance contains all states when $a=\fence$. 
For an action $a\in Wr\cup Rd$, $\vmax_{PP}(t,a)$ contains all states $\sigma$ with maximal timestamp for thread $t$ and variable $x=var(a)$. 
This either means all write buffers for $x$ are empty or the last timestamp in $\sigma.wb^{t,x}$ is larger than any other timestamp in the write buffers for $x$. Also note that $FL_{PP}$ fulfils the properties of Definition \ref{def:beta} and therefore $FL_{PP}$ is at least a subset of the view-preserving simulation $\beta$.

\begin{theorem}\label{th:psoaxioms}
PSO satisfies the axioms of {\sc Core}, {\sc SharedVars} and {\sc Fences}, but not the axiom of {\sc MsgPassing}.
\end{theorem}

For the proofs of the satisfied axioms, see the appendix, where we show that PPSO fulfils them. Because of Corollary \ref{cor:treq} the axioms then also hold for PSO. 
We can show that MP does not hold for PPSO by looking at Figure \ref{fig:mp}. 
A possible outcome of the example is $r_1=1\land r_2=0$ (see Section \ref{sec:pso}). Let $a_w=wr(y,1)$, $a_r=rd(y,r_1,1)$, $b=wr(x,1)$, $t=1$ and $t'=2$. We choose $\sigma$ to be the state after the write $x:=1$ in thread 1, such that $\sigma.wb^{1,x}=\langle (1,2) \rangle$. Hence $\sigma\in \vmax_{PP}(1,b)$ and since $\sigma\notin \dom(T_{PP}(2,a_r))$, we also get $\sigma\in \wlp(T_{PP}(2,a_r),\vmax_{PP}(2,b))$. Now we need to show, that $\sigma\notin \wlp(T_{PP}(1, a_w), \wlp(T_{PP}(2, a_r), \vmax (2, b)))$. For that let $\sigma'$ be a state after $y:=1$, such that $\sigma'.wb^{1,x}=\langle (1,2) \rangle$ and $\sigma'.wb^{1,y}=\langle (1,1) \rangle$. After a flush of $wb^{1,y}$ we get to a state $\sigma''$ for which $\sigma''.s(y)=1$ and still $\sigma''.wb^{1,x}=\langle (1,2) \rangle$. Then $(\sigma, \sigma')\in T_{PP}(1,a_w)$ and $(\sigma',\sigma'')\in T_{PP}=(2,a_r)$, but $\sigma''\notin \vmax_{PP}(2,b)$. Hence, $\sigma\notin \wlp(T_{PP}(1, a_w), \wlp(T_{PP}(2, a_r), \vmax_{PP}(2, b)))$ and therefore PPSO and PSO do not satisfy MP. 

\section{Example Proof}
In this section we want to apply this form of axiomatic reasoning to prove the correctness of the message passing with fence litmus test (see Fig.~\ref{fig:mp-fence}). For that we use Owicki-Gries reasoning for concurrent programs \cite{DBLP:journals/acta/OwickiG76}. Like \cite{TOCL} we prove the holding of Hoare-triples \cite{DBLP:journals/cacm/Hoare69} of the form $\{P\}com_t\{Q\}$ by showing that $$P\subseteq\wlp(T(t,a),Q)$$ holds for all actions $a$ belonging to $com_t$. Here $com_t$ is the program command of a thread $t$ and we call $P$ the {\em pre-assertion} and $Q$ the {\em post-assertion}. 
These assertions occur in proof outlines of programs as in Owicki-Gries reasoning \cite{DBLP:journals/acta/OwickiG76}. The assertion above a command is its pre-assertion and the one below its post-assertion. To check if a proof outline is valid we differentiate between local and global correctness (interference-freedom). A proof outline is valid when every thread is locally correct and the proof outline itself is globally correct. 

\begin{definition}[\cite{TOCL}]\\
\label{def:correct}
 A thread $t$ is {\em locally correct} in a proof outline if $\{P\}com_t\{Q\}$ holds for every program command $com$ in $t$ with pre-assertion $P$ and post-assertion $Q$.  
 
 A proof outline is {\em globally correct} if for every pair of threads $t, t'$   $\{R\cap P\}com_{t'}\{R\}$ holds for every assertion $R$ in the proof outline of $t$ and command $com$ with pre-assertion $P$ in thread $t'$.
\end{definition}

\begin{figure}[t]
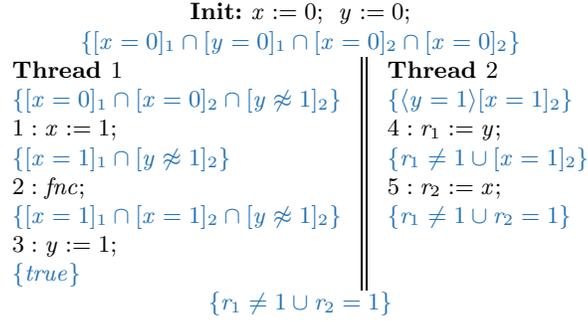

    \begin{center} 
\begin{tabular}[b]{@{}c@{}} 
   {\bf Init:} $x:=0;$ $y:=0$;   \\ 
   ${\color{mycolor} \{[x=0]_1 \cap [y=0]_1 \cap [x=0]_2 \cap [x=0]_2\} }$ \\ 
  $\begin{array}{l@{\ \ }||@{\ \ }l}
    \begin{array}[t]{l}
     \textbf{Thread } 1 
     \\
     {\color{mycolor} \{[x=0]_1 \cap [x=0]_2 \cap [y \not \approx 1]_2\} } \\ 
     1: x:=1 ;
     \\
    {\color{mycolor} \{[x=1]_1 \cap [y \not \approx 1]_2\} } \\
     2: fnc ;
     \\ 
    {\color{mycolor} \{[x=1]_1 \cap [x=1]_2 \cap [y \not \approx 1]_2\} } \\
     3: y:=1 ;
     \\
    {\color{mycolor} \{true\}  }
     \end{array}
    & 
    \begin{array}[t]{l}
     \textbf{Thread } 2 
     \\
    {\color{mycolor} \{\langle y = 1\rangle [x = 1]_2 \} } \\
     4: r_1:=y ;
     \\
     {\color{mycolor} \{r_1 \neq 1 \cup [x = 1]_2 \} }\\
    5: r_2:=x ;
     \\
     {\color{mycolor} \{r_1 \neq 1 \cup r_2 = 1 \} } 
     \end{array}
   \end{array}$ \\
   ${\color{mycolor} \{r_1 \neq 1 \cup r_2 = 1\} }$ \\
   \end{tabular} 
\end{center}
\vspace*{-0.5cm}
    \caption{Proof outline for MP with fence}
    \vspace*{-0.3cm}
    \label{fig:exampleproof}
\end{figure}

Next we look at the MP with fence example. In the proof outline, we use a type of assertion additionally to the ones defined in Def.~\ref{def:assertions}. For $t\in\Tid$, $x,y\in \VarG$ and $u,v\in \Val$ we call 
\begin{align*}
    \langle y=u \rangle [x=v]_t=\wlp(T(t,rd(y,r,u)),[x=v]_t)
\end{align*}
the {\em conditional observation assertion}. It holds for all states in which the synced value assertion $[x=v]_t$ holds, if $t$ can read the value $u$ for $y$ first. 
Since the assertion contains only $\beta$-stable predicates, it is $\beta$-stable itself.
The reasoning technique only uses the axioms holding for PSO. 
The proof outline given thus holds for all memory models satisfying these axioms. 

\begin{lemma}\label{th:proofoutline}
The proof outline in Figure~\ref{fig:exampleproof} is valid under {\sc Fences}.
\end{lemma}

To prove this lemma, we need to show local and global correctness of the proof outline. We illustrate how the proof looks like in detail by proving a Hoare-triple for the global correctness of thread 2. For that we choose the assertion $\purple{\langle y = 1\rangle [x = 1]_2}$ from the proof outline of thread 2 and the command $y:=1$ of thread 1 with its pre-assertion $\teal{[x=1]_1 \cap [x=1]_2 \cap [y \not \approx 1]_2}$: 
\begin{align*}
    \{ \purple{\langle y = 1\rangle [x = 1]_2} \cap \teal{[x=1]_1 \cap [x=1]_2 \cap [y \not \approx 1]_2} \}  y:=_11  \{ \purple{\langle y = 1\rangle [x = 1]_2} \} 
\end{align*}
This means we show that 
\begin{align*}
    &\purple{\wlp(T(2,rd(y,r_1,1)),[x = 1]_2)} \cap \teal{[x=1]_1 \cap [x=1]_2 \cap [y \not \approx 1]_2} 
    \\&\subseteq \wlp(T(1,wr(y,1)),\purple{\wlp(T(2,rd(y,r_1,1)),[x = 1]_2)})
\end{align*}
Since $\wlp(T(2,rd(y,r_1,1)),[x = 1]_2) \cap [x=1]_1 \cap [x=1]_2 \cap [y \not \approx 1]_2 \subseteq [x=1]_2 $, $[x=1]_2 \subseteq \wlp(T(2,rd(y,r_1,1)),[x = 1]_2)$ (see Lemma \ref{lem:readstable}) and the monotonicity rule from Figure \ref{fig:wlp}, we only need to prove that
\begin{align*}
    [x=1]_2 \subseteq \wlp(T(1,wr(y,1)),[x = 1]_2)
\end{align*}
We divide this into two parts: (i) $[x \equiv 1]_2 \subseteq \wlp(T(1,wr(y,1)),[x \equiv 1]_2)$ and (ii) $x_{\uparrow 2} \subseteq \wlp(T(1,wr(y,1)),x_{\uparrow 2})$. Then with the conjunctivity rule from Figure \ref{fig:wlp} the Hoare-triple is proven. 

For (i), we need to show for every value $v\neq 1$:
\begin{align*}
    \dis(T(2,rd(x,r_2,v))) \subseteq \wlp(T(1,wr(y,1)),\dis(2,rd(x,r_2,v)))
\end{align*}
Using the rules of Figure \ref{fig:wlp} and the definition of $\dis$ we get
\begin{align*}
    \dis(T(2,rd(x,r_2,v))) 
    &\subseteq \wlp(T(2,rd(x,r_2,v)),\emptyset)\\
    &\subseteq \wlp(T(2,rd(x,r_2,v)),\wlp(\interf(1,wr(y,1)),\emptyset))\\
    &\subseteq \wlp(T(2,rd(x,r_2,v))\fcmp \interf(1,wr(y,1)),\emptyset)
\end{align*}
After applying \textbf{SV1}, we get
\begin{align*}
    &\wlp(T(2,rd(x,r_2,v))\fcmp \interf(1,wr(y,1)),\emptyset)\\
    &\subseteq \wlp(\interf(1,wr(y,1))\fcmp T(2,rd(x,r_2,v)),\emptyset)\\
    &\subseteq \wlp(\interf(1,wr(y,1)),\wlp(T(2,rd(x,r_2,v)),\emptyset))\\
    &\subseteq \wlp(\interf(1,wr(y,1)),\dis(T(2,rd(x,r_2,v))))
\end{align*}
Now (i) follows from Lemma \ref{lem:overapprox} and the conjunctivity rule.

For (ii), we use \textbf{SV2} and get for every action $a\in\Act_{|x}$
\begin{align*}
    \vmax(2,a) \subseteq \wlp(T(1,wr(y,1)),\vmax(2,a))
\end{align*}
Hence (ii) follows from the conjunctivity rule.
\section{Related Work} \label{sec:related} 
The work closest to us is that of Doherty et al.~\cite{TOCL} who gave a generic approach to  Owicki-Gries based axiomatic reasoning and applied that to a number of weak memory models (SC, C11 RAR and TSO). Since we show where PSO fits in their axiomatic hierarchy, our program semantics and the axiomatic reasoning are based on their work. We supplement that by giving a PSO specific semantics, showing which axioms hold for PSO and using them on a correctness proof of the MP with fence litmus test.

Other Owicki-Gries related approaches on reasoning for weak memory models were made by \cite{DBLP:conf/esop/BilaDLRW22,ecoop20,DBLP:journals/corr/abs-2004-02983,DBLP:conf/icalp/LahavV15}.
While Bila et al.~\cite{DBLP:conf/esop/BilaDLRW22} present an Owicki-Gries based logic for the persistent TSO memory model, 
Dalvandi et al.~\cite{ecoop20,DBLP:journals/corr/abs-2004-02983} and Lahav and Vafeiadis \cite{DBLP:conf/icalp/LahavV15} looked at the C11 memory model. None of them introduce a generic proof calculus or a PSO specific one. 

There are however other approaches regarding verification specific to PSO \cite{DBLP:journals/acta/AbdullaAAJLS17,DBLP:conf/esop/AlglaveKNT13,DBLP:conf/vmcai/DanMVY15,DBLP:journals/monet/XiaoZXV22}. 
Abdulla et al.~\cite{DBLP:journals/acta/AbdullaAAJLS17} for instance introduce a technique for stateless model checking under TSO and PSO. It is based on chronological traces, which define a partial order relation on program executions. 
Alglave et al.~\cite{DBLP:conf/esop/AlglaveKNT13} present a transformation technique to verify programs under weak memory model by using SC tools. For different memory models (TSO, PSO, POWER,...) they define an abstract machine which is equivalent to its axiomatic semantics. They then identify so-called unsafe pairs in the program and linearise it. The SC analyser then verifies the new program. 
Dan et al.~\cite{DBLP:conf/vmcai/DanMVY15} abstract store buffers of TSO and PSO to bounded arrays. For both of these models, they are able to translate a given program into one with over-approximately the same behaviour on SC and then also use SC Tools for its verification. 
Xiao et al.~\cite{DBLP:journals/monet/XiaoZXV22} formalise the PSO memory model with the process algebra Communicating Sequential Processes and use this formalisation for a model checker. 
Some of these techniques do not only work for PSO. Still, they are not generic since they cannot be easily applied to a different memory model.

More generic approaches were made by \cite{DBLP:conf/popl/AlglaveC17,DBLP:conf/fmcad/LeonFHM18,DBLP:conf/cav/GavrilenkoLFHM19,DBLP:conf/pldi/Kokologiannakis19}.
Alglave and Cousot \cite{DBLP:conf/popl/AlglaveC17} present an invariance proof method which shows that a given program is correct w.r.t.~a given memory model and an invariant specification of that program. It does so by first proving that a so-called communication specification is sufficient for the program's invariant. 
If a memory model guarantees the communication, the program is correct under that model. 
Their method differs from the one we are using. Since every memory model needs its own communication specification, each step of the long proof has to be redone for a different model. This makes their proofs more complex. 
The other work focusses on model checking.
Ponce de Leon et al.~\cite{DBLP:conf/fmcad/LeonFHM18} and Gavrilenko et al.~\cite{DBLP:conf/cav/GavrilenkoLFHM19} present generic bounded model checkers which translate a given program under a given memory model into an SMT formula. They are generic because their input contains not only the program but also the memory model, formalised in CAT as a set of relations.
Finally, Kokologiannakis et al.~\cite{DBLP:conf/pldi/Kokologiannakis19} developed a generic model checker that transforms a given program into an execution graph to check its correctness under a given memory model with an axiomatic semantics.

\section{Conclusion} \label{sec:conclusion}

In the paper, we have shown the weak memory model PSO to instantiate the axioms of~\cite{TOCL} for all but one axiom, thereby enabling memory-model independent reasoning. We have exemplified the reasoning technique on a litmus test for message passing achieved via the insertion of a fence. 

As future work, we see the lifting of these low-level axioms to the level of view-based assertions as to provide a more abstract level for reasoning. We also plan to look at other weak memory models to check whether they fulfil the axioms.  

\bibliographystyle{splncs04}
\bibliography{references}

\appendix 
\section{Proofs}
In the appendix we prove the lemma and theorems of Section \ref{sec:ppso}. In the first part we show that $TS_{PP}$ and $TS_{PSO}$ are trace equivalent. Then in the second part we use this to show our main result, that PSO satisfies the axioms of \textsc{Core}, \textsc{SharedVars} and \textsc{Fences}.
\subsection{Forward and Backward Simulations}
Let $TS_{PP}=(\Act,\Sigma_{PP},I_{PP},T_{PP})$ and $TS_{PSO}=(\Act,\Sigma_{PSO},I_{PSO},T_{PSO})$.
\subsubsection{Proof of Theorem \ref{th:fsim}.} 
Let $F\subseteq\Sigma_{PP}\times\Sigma_{PSO}$ where 
\begin{align*}
(\sigma_1,\sigma_2)\in F : \Leftrightarrow s_1=s_2 \wedge \forall t\in \Tid \text{ } \forall x\in \VarG  : wb_2^{t,x}= wb_1^{t,x}.val
\end{align*}
with $\sigma_1=(s_1,wb_1)$ and $\sigma_2=(s_2,wb_2)$. 
We want to prove that there exists a forward simulation from $TS_{PP}$ to $TS_{PSO}$. Therefore, we look at the initial states first. Let $\sigma_{01}=(s_{01},wb_{01})\in I_{PP}$. We need to show that there exists a $\sigma_{02}\in I_{PSO}$ with $(\sigma_{01},\sigma_{02})\in F$. Let $\sigma_{01}=(s_{01},wb_{01})\in I_{PP}$ and $\sigma_{02}=(s_{02},wb_{02})$ with $s_{02}=s_{01}$ and $wb_{02}^{t,x}=wb_{01}^{t,x}.val$. Then $\sigma_{02}\in I_{PP}$ and  $(\sigma_{01},\sigma_{02})\in F$.
\\\\Now we show that 
\textbf{\begin{align*}
\forall (\sigma_1,\sigma_2)\in F \text{ } \forall a\in \Act: (\sigma_1,\sigma'_1)\in T_{PP}(t,a)\\  \Rightarrow \exists \sigma'_2 : (\sigma_2,\sigma'_2)\in T_{PSO}(t,a) \wedge (\sigma'_1,\sigma'_2)\in F.
\end{align*}} 
Let $(\sigma_1,\sigma_2)\in F$ where $\sigma_1=(s_1,wb_1)\in \Sigma_{PP}$ and $\sigma_2=(s_2,wb_2)\in\Sigma_{PSO}$. Let $\sigma_1'=(s_1',wb_1')\in\Sigma_{PP}$ with $(\sigma_1,\sigma_1')\in T_{PP}(t,a)$. If $(\sigma_1,\sigma_1'')\in FL_{PP}$ and $(\sigma_2,\sigma_2'')\in FL_{PSO}$, both flush rules change $s$ in the same way and delete the first entry of $wb_1^{t,x}$ and $wb_2^{t,x}$, then $(\sigma_1'',\sigma_2'')\in F$. Therefore, we only need to show that $(\sigma_1',\sigma_2')\in F$ for $\sigma_1''\overset{a,t}{\leadsto}_{PP}\sigma_1'$ and $\sigma_2''\overset{a,t}{\leadsto}_{PSO}\sigma_2'$:
\begin{itemize}
    \item Case 1: $a=rd(x,r,v)$. $\overset{a,t}{\leadsto}_{PP}$ and $\overset{a,t}{\leadsto}_{PSO}$ do not change $\sigma_1''$ or $\sigma_2''$. Hence  $(\sigma'_1,\sigma'_2)=(\sigma''_1,\sigma''_2)\in F$.
    \item Case 2: $a=fence$. If a state $\sigma_1''$ satisfies the conditions for PP-Fence, then $\sigma_2''$ does for PSO-Fence. Both states will not change after a fence action, hence $(\sigma'_1,\sigma'_2)=(\sigma''_1,\sigma''_2)\in F$.
\end{itemize}

\subsubsection{Proof of Lemma \ref{lem:bsim}.}
Let $\sigma\in\Sigma_{PP}$ and $\sigma_0\in I_{PP}$ with $\sigma_0 \overset{\alpha}{\Rightarrow}\sigma$. Then there exist $a_1,...,a_k\in \Act\cup\{flush\}$ and  $\sigma_1,...,\sigma_{k-1}\in \Sigma_{PP}$ such that 
$$\sigma_0 \overset{a_1}{\leadsto}_{PP} \sigma_1\overset{a_2}{\leadsto}_{PP}...\overset{a_{k-1}}{\leadsto}_{PP}\sigma_{k-1}\overset{a_k}{\leadsto}_{PP}\sigma.$$
Since $\sigma_0\in I_{PP}$, $\sigma_0.wb^{t',x'}$ is empty for every $t'\in\Tid$ and $x'\in \VarG$. Write buffers only change after a write or flush action. 
Let $\sigma_i$ be the first state of the sequence with a non empty write buffer $\sigma_i.wb^{t_i,x_i}$ for $i\in \{1,...,k-1\}$, $t_i\in \Tid$ and $x_i\in \VarG$. Then $a_i=wr(x_i,v_i)$ has to be a write action and $\sigma_i.wb^{t_i,x_i}=\langle(v_i,q_i)\rangle$ for a $q_i\in \mathbb{Q}$. Hence $\sigma_i$ fulfils the invariant.
Let $\sigma_{j-1}\in\Sigma_{PP}$ with $i+1\leq j< k$, $\sigma_{j-1}.wb\neq\sigma_j.wb$ and $\sigma_{j-1}$, and every state in the sequence before $\sigma_{j-1}$, fulfil the invariant. Now we show that $\sigma_j$ fulfills the invariant for every action $a_j$. Since read and fence actions do not change $\sigma_{j-1}$, we only need to look at write and flush actions. Then every state in our sequence, $\sigma$ in particular, fulfils the invariant and the lemma is proven. 
\begin{itemize}
    \item Case 1: $a_j=wr(x_j,v_j)$. In this case $fresh_{\sigma_{j-1}}(t_j,x_j,q_j)=true$, which means 
    \begin{align*}
        (\forall t' \in \Tid \text{ } \forall x' \in \VarG : (\cdot,q_j) \notin \sigma_{j-1} .wb^{t',x'})\\ \wedge (\forall (\cdot,q') \in \sigma_{j-1} .wb^{x_j,t_j} : q_j>q').
    \end{align*}
    Hence, we get for $\sigma_j$
    \begin{align*}
        \forall t' \in \Tid \text{ } \forall x' \in \VarG,t'&\neq t_j \lor x'\neq x_j: (\cdot,q_j) \notin \sigma_j .wb^{t',x'}\\ \land\forall (\cdot,q') \in \sigma_j .wb^{x_j,t_j},(\cdot,&q')\neq last(\sigma_j.wb^{t_j,x_j}): q_j>q'.
    \end{align*}
    Since $a_j$ only changes $wb^{t_j,x_j}$ and since $\sigma_{j-1}$ already fulfils the invariant, so does $\sigma_j$.
    \item Case 2: $a_j=flush$. Then $\sigma_j.wb^{t_j,x_j}$ has one entry less than $\sigma_{j-1}.wb^{t_j,x_j}$. If $\sigma_{j-1}.wb^{t_j,x_j}$ had only one entry, then $\sigma_j.wb^{t_j,x_j}$ is empty and $\sigma_j$ satisfies the invariant. If $\sigma_{j-1}.wb^{t_j,x_j}$ has more than one entry, we get $ts(last(\sigma_{j}.wb^{t_j,x_j}))\!=\!ts(last(\sigma_{j-1}.wb^{t_j,x_j}))$ and $\{(\cdot,q') \in \sigma_j .wb^{x_j,t_j} | (\cdot,q')\neq last(\sigma_j.wb^{t_j,x_j})\}\subseteq\{(\cdot,q') \in \sigma_{j-1} .wb^{x_j,t_j} | (\cdot,q')\neq last(\sigma_{j-1}.wb^{t_j,x_j})\}$. Hence $\sigma_j$ fulfils the invariant. 
\end{itemize}

\subsubsection{Proof of Theorem \ref{th:bsim}.} 
Let $B\subseteq\Sigma_{PSO}\times\Sigma_{PP}$ where 
\begin{align*}
(\sigma_1,\sigma_2)\in B : \Leftrightarrow s_1=s_2 \wedge \forall t\in \Tid \text{ } \forall x\in \VarG  : wb_1^{t,x}= wb_2^{t,x}.val.
\end{align*}
This relation is total on $\Sigma_{PSO}$.
We want to prove that there exists a backward simulation from $TS_{PSO}$ to $TS_{PP}$. 
Therefore, we look at the initial states first. Let $\sigma_{01}\in I_{PSO}$. We need to show that for every $\sigma_{02}\in \Sigma_{PP}$ with $(\sigma_{01},\sigma_{02})\in B$ we get $\sigma_{02}\in I_{PP}$. Let $\sigma_{01}=(s_{01},wb_{01})$. Then $wb_{01}^{t,x}=\langle \text{ } \rangle$ for every $t\in \Tid$ and $x\in \VarG$. Hence, for every $\sigma_{02}$ with $(\sigma_{01},\sigma_{02})\in B$ all write buffers are empty, and therefore $\sigma_{02}\in I_{PP}$. 
Now we show that for every $(\sigma'_1,\sigma'_2)\in B$ and every action $a$
\begin{align*}
    (\sigma_1, \sigma'_1)\in T_{PSO}(t,a)  \Rightarrow \exists \sigma_2 : (\sigma_2,\sigma'_2)\in T_{PP}(t,a) \wedge (\sigma_1,\sigma_2)\in B
\end{align*}
Let $(\sigma'_1,\sigma'_2)\in B$ for $\sigma'_1=(s'_1,wb'_1)\in\Sigma_{PSO}$ and $\sigma'_2=(s'_2,wb'_2)\in\Sigma_{PP}$. Let $\sigma_1=(s_1,wb_1)\in\Sigma_{PSO}$ with $(\sigma_1,\sigma_1')\in T_{PSO}(t,a)$. Then there exists a state $\sigma_1''=(s_1'',wb_1'')$ with $(\sigma_1,\sigma_1'')\in FL_{PSO}$ and $\sigma_1'' \overset{a,t}{\leadsto}_{PSO}\sigma_1'$. First we show that for every $a\in\Act$ exists a $\sigma_2''\in\Sigma_{PP}$ with $\sigma_2'' \overset{a,t}{\leadsto}_{PP}\sigma_2'$ and $(\sigma_1'',\sigma_2'')\in B$.

\begin{itemize}
    \item Case 1: $a=rd(x,r,v)$. The read actions do not change $\sigma_1''$ and $\sigma_2''$. Hence $\sigma_1''=\sigma'_1$ and $\sigma_2''=\sigma'_2$ and therefore $(\sigma_1'',\sigma_2'')\in B$.
    \item Case 2: $a=fence$. If $\sigma_1''$ holds the condition for PSO-fence then $\sigma_2''$ holds the one for PP-fence. The fence actions do not change $\sigma_1''$ and $\sigma_2''$. Hence $\sigma_1''=\sigma'_1$ and $\sigma_2''=\sigma'_2$ and therefore $(\sigma_1'',\sigma_2'')\in B$.
    \item Case 3: $a\!=\!wr(x,v)$. In this case $\sigma_1'.wb^{t,x}\!=\!\sigma_1''.wb^{t,x}\cdot\langle v\rangle$. Since $(\sigma_1',\sigma_2')\!\in\! B$, we have $last(\sigma_1'.wb^{t,x})\!=\!v\!=\!val(last(\sigma_2'.wb^{t,x}))$. Let $q\!=\!ts(last(\sigma_2'.wb^{t,x}))$. Now we can apply Lemma \ref{lem:bsim} and get 
    $$\exists\sigma_2''\in\Sigma_{PP}:fresh_{\sigma_2''}(t,x,q)=true\land\sigma_2'' \overset{a,t}{\leadsto}_{PP}\sigma_2'$$
    Since both write actions change $wb^{t,x}$ by adding a new entry, we get $(\sigma_1''\!,\sigma_2'')\!\in\! B$.
\end{itemize}
Now we need to show that there exists $\sigma_2\in\Sigma_{PP}$ with $(\sigma_2,\sigma_2'')\in FL_{PP}$ and $(\sigma_1,\sigma_2)\in B$. For $\sigma_1=\sigma_1''$, we set $\sigma_2=\sigma_2''$. If $\sigma_1\neq\sigma_1''$, let $n$ be the number of flush actions between $\sigma_1$ and $\sigma_1''$: 
$$\sigma_1\!=\!\sigma_{1,n}\!\overset{flush,t_n}{\leadsto}_{PSO}\sigma_{1,n-1}\!\overset{flush,t_{n-1}}{\leadsto}_{PSO}\ldots\!\overset{flush,t_2}{\leadsto}_{PSO}\sigma_{1,1}\!\overset{flush,t_1}{\leadsto}_{PSO}\sigma_{1,0}\!=\!\sigma_1''$$
Here $\sigma_{1,1},\ldots,\sigma_{1,n-1}\!\in\!\Sigma_{PSO}$ and for $1\!\leq\! i\!\leq\! n$ $wb^{t_i,x_i}$ is flushed by $\overset{flush,t_i}{\leadsto}_{PSO}$ with $x_i\!\in\! \VarG$. 
Every flush action $\overset{flush,t_i}{\leadsto}_{PSO}$ deletes the first entry of $wb^{t_i,x_i}$ and changes the value $s(x_i)$. 
Let $\sigma_{2,0}\!=\!\sigma_2''$ and $\sigma_{2,i}\!\in\!\Sigma_{PP}$ with $\sigma_{2,i}.wb^{t,x}\!=\!\sigma_{2,i-1}.wb^{t,x}$ for every $\sigma_{2,i}.wb^{t,x}\neq \sigma_{2,i}.wb^{t_i,x_i}$. 
We set $\sigma_{2,i}.wb^{t_i,x_i}(j+1)=\sigma_{2,i_1}.wb^{t_i,x_i}(j)$ for every entry $\sigma_{2,i_1}.wb^{t_i,x_i}(j)$ and $\sigma_{2,i}.wb^{t_i,x_i}(0)\!=\!(\sigma_{2,i-1}.s(x_i),q)$ with a timestamp $q$ that is smaller than every timestamp in every write buffer of $\sigma_{2,i-1}$. 
Then 
$$\sigma_{2,n}\overset{flush,t_n}{\leadsto}_{PP}\sigma_{2,n-1}\overset{flush,t_{n-1}}{\leadsto}_{PP}\ldots\overset{flush,t_2}{\leadsto}_{PP}\sigma_{2,1}\overset{flush,t_1}{\leadsto}_{PP}\sigma_{2,0}=\sigma_2''$$
and $(\sigma_{1,i},\sigma_{2,i})\in B$ for every $i$. Hence if we se $\sigma_2=\sigma_{2,0}$, we get $(\sigma_2,\sigma_2'')\in FL_{PP}$ and $(\sigma_1,\sigma_2)\in B$.
\subsection{Axiom Proofs}
In this section we prove Theorem \ref{th:psoaxioms} by showing that PPSO fulfils every axiom of the levels \textsc{Core}, \textsc{SharedVars} and \textsc{Fences}.

\subsubsection{{\sc Core} Axioms}\,
\\\\\textbf{Proof of C1.}     
    \begin{itemize}
        \item $a=\fence$. Then $\vmax_{PP}(t,a)=\Sigma_{PP}\supseteq I_{PP}$.
        \item $a\in Wr\cup Rd$ with $\var(a)=x$. In that case 
	        \begin{align*}
	        \vmax_{PP}(t,a)
	        &=maxTS(t,x)\\
	        &\supseteq \{\sigma\in \Sigma_{PP}|\forall t'\in \Tid:\sigma.wb^{t',x}=\langle\,\,\rangle\}\\
	        &\supseteq I_{PP}.
	        \end{align*}
    \end{itemize}
\textbf{Proof of C2.}
    \begin{itemize}
        \item $a=\fence$ or $a=rd(x,r,v)$. Then $T_{PP}(t,a)=FL =T_{PP}(t',a)$. Let $\sigma\in \vmax_{PP}(t,a)$ and $\sigma'\in\Sigma_{PP}$ with $(\sigma,\sigma')\in T_{PP}(t,a)$. For $\tau=\sigma'$ we get $(\sigma'\tau)\in \beta$ and $(\sigma,\tau)\in T_{PP}(t',a)\subseteq T_{PP}(t',a)\fcmp \beta$.
        \item $a=wr(x,v)$. Let $\sigma\in \vmax_{PP}(t,a)=maxTS(t,x)$ and $(\sigma,\sigma')\in T_{PP}(t,a)$. Let $(\sigma',\tau)\in FL$ where we flush at least as often as we need for $\sigma'.wb^{t,x}$ to become empty. Hence $(\sigma',\tau)\in\beta$ and $\tau.s(x)=v$. Since $\sigma\in maxTS(t,x)$ we get $\tau.wb^{t'',x}=\langle\ \rangle$ for all $t''\in \Tid$. Therefore $(\sigma,\tau)\in T_{PP}(t',a)\fcmp\beta$.
    \end{itemize}
\textbf{Proof of C3.}
    \begin{itemize}
        \item $a=\fence$ or $a\in Rd$. Then $\interf_{PP}(t,a)=id$ and therefore $T_{PP}(t,a)=FL\subseteq \beta\subseteq\beta\fcmp \interf_{PP}(t,a)\fcmp\beta$.
        \item $a=wr(x,v)$. In this case $\beta\fcmp \interf_{PP}(t,a)\fcmp\beta\supseteq FL\fcmp T_{PP}(t,a)\fcmp FL\supseteq T_{PP}(t,a)$.
   \end{itemize}
\textbf{Proof of C4.}
    \begin{itemize}
        \item $b=\fence$ or $b\in Rd$. Then $\interf_{PP}(t,b)\, =\, id$ and hence $\wlp(\interf_{PP}(t,b),\\\vmax_{PP}(t,a))=\vmax_{PP}(t,a)$.
        \item $b=wr(x,v)$. If $a=\fence$ then $\vmax_{PP}(t,a)=\Sigma_{PP}$. Hence $\vmax_{PP}(t,a)\\= \wlp(\interf_{PP}(t,b),\vmax_{PP}(t,a))$. For $a\in Wr\cup Rd$ let $\sigma\in \vmax_{PP}(t,a)$. Since $\interf_{PP}(t,b)$ can only add an entry to $wb^{t,x}$, we get $\sigma'\in \vmax_{PP}(t,a)$ and therefore $\sigma\in \wlp(\interf_{PP}(t,b),\vmax_{PP}(t,a))$.
    \end{itemize}

\subsubsection{{\sc SharedVars} Axioms}\,
\\\\\textbf{Proof of SV1.}
    \begin{itemize}
        \item $b=\fence$ or $b\in Rd$. Then $\interf_{PP}(t,b)\, =\, id$ and therefore $\interf_{PP}(t',b)\fcmp T_{PP}(t,a)\subseteq T_{PP}(t,a)\fcmp \interf_{PP}(t',b)$.
        \item $b=wr(x,v)$. If $a\notin Wr$ then $T_{PP}(t,a)=FL$. Hence, we only look at $a=wr(y,u)$. Since $\var(a)\neq \var(b)$ we can switch the two writes.
    \end{itemize}
\textbf{Proof of SV2.}
    \begin{itemize}
        \item $b=\fence$ or $b\in Rd$. Then $\interf_{PP}(t',b)\, =\, id$ and therefore\\ $\wlp(\interf_{PP}(t',b),\vmax_{PP}(t,a))=\vmax_{PP}(t,a)$.
        \item $b=wr(x,v)$. If $a=\fence$ then $\vmax_{PP}(t,a)=\Sigma_{PP}$. Hence $\vmax_{PP}(t,a)\\= \wlp(\interf_{PP}(t',b),\vmax_{PP}(t,a))$. For $a\in Wr\cup Rd$ let $\sigma\in \vmax_{PP}(t,a)$. Since $\interf_{PP}(t',b)$ can only add an entry to $wb^{t',x}$ and $\var(a)\neq x$ we get $\sigma'\in \vmax_{PP}(t,a)$ and therefore $\sigma\in \wlp(\interf_{PP}(t,b),\vmax_{PP}(t,a))$.
    \end{itemize}
\textbf{Proof of RW1.}
Since $a_r$ is a read, we get 
    \begin{align*}
        \interf_{PP}(t',a_w) \fcmp T_{PP}(t,a_r) 
        &= T_{PP}(t',a_w) \fcmp FL \\
        &= FL \fcmp \overset{a_w,t'}{\leadsto}_{PP} \fcmp FL \\
        &= FL \fcmp T_{PP}(t',a_w) \fcmp FL \\
        &= T_{PP}(t,a_r) \fcmp \interf_{PP}(t',a_w).
    \end{align*}
\textbf{Proof of RW2.}
Since $a_r$ is a read $\interf_{PP}(t',a_r)=id$. Hence $$\interf_{PP}(t',a_r)\fcmp T_{PP}(t,a)= T_{PP}(t,a)\fcmp \interf_{PP}(t',a_r).$$
\textbf{Proof of RW3.}
Since $a_r$ is a read $\interf_{PP}(t',a_r)=id$. Hence $$\vmax_{PP}(t,a)\subseteq \wlp(\interf_{PP}(t',a_r),\vmax_{PP}(t,a)).$$
\textbf{Proof of RW4.}
We have 
\begin{align*}
    \dom(T_{PP}(t,Rd_{|x})) &= \{ \sigma | \exists v \in \Val: \sigma.s(x)=v \lor wbVal_{\sigma}(t,x)=v\}\\ &= \Sigma_{PP}.
\end{align*}
\textbf{Proof of RW5.}
Let $\sigma\in\Sigma_{PP}$ with $(\sigma,\sigma')\!\in\! T_{PP}(t,a_w)$. Then $wbVal_{\sigma'}(t,x)\!=\!v$ or $\sigma'.s(x)=v$. Hence $\sigma'\in \dom(T_{PP}(t,Rd_{|x}[v]))$ and therefore $$\Sigma_{PP}\subseteq \wlp(T_{PP}(t,a_w),\dom(T_{PP}(t,Rd_{|x}[v]))).$$
\textbf{Proof of RW6.}
Since $x_{\uparrow t} = \bigcap_{a\in \Act_{PP|x}} \vmax_{PP}(t,a) = maxTS(t,x)$ and 
\begin{align*}
    \bigcup_{v\in \Val}[x\equiv v]_t
    &= \bigcup_{v\in \Val} \bigcap_{u \neq v} [x\napprox v]_t \\
    &= \bigcup_{v\in \Val} \bigcap_{u \neq v} \dis(T_{PP}(t,Rd_{|x}[u])) \\
    &= \bigcup_{v\in \Val} \{\sigma | \sigma.s(x)=v  \lor wbVal_{\sigma}(t,x)=v\} \\
    &= \Sigma_{PP}
\end{align*}
we get $x_{\uparrow t}\subseteq \bigcup_{v\in \Val}[x\equiv v]_t$.
\\\\\textbf{Proof of RW7.}
Let $a_w=wr(x,v)$ and $a_r=rd(x,r,v)$. 
For $\sigma \in maxTS(t,x)$ if $\sigma.wb^{t,x} \neq \langle\,\rangle$, then $\sigma \notin \dis(T_{PP}(t',a_r))$. Hence, for every $\sigma \in \vmax_{PP}(t,a_w)\cap \dis(T_{PP}(t',a_r))$ we get $\sigma.wb^{t,x}=\langle\,\rangle$. Then $\sigma'.s(x)\neq v$ for all $(\sigma,\sigma')\in FL$. Let $(\sigma,\sigma')\in T_{PP}(t,a_w)$. Then $\sigma'.wb^{t,x}=\langle (v,q) \rangle$ and since $t\neq t'$ we get $\sigma''.wb^{t,x}=\langle\,\rangle$ for $(\sigma',\sigma'')\in T_{PP}(t',a_r)$. Therefore, $\sigma'' \in \vmax_{PP}(t',a)$ and $\sigma \in \wlp(T_{PP}(t,a_w),\wlp(T_{PP}(t',a_r),\vmax_{PP}(t',a)))$.

\subsubsection{{\sc Fences} Axiom}\,
\\\\\textbf{Proof of FNC.}
If $a=\fence$, $\vmax_{PP}(t,a)=\Sigma_{PP}=\vmax_{PP}(t',a)$ and therefore $\vmax_{PP}(t,a)= \wlp(T_{PP}(t,\fence),\vmax_{PP}(t',a))$. 
Let $a\in Wr\cup Rd$ and $\var(a)=x$. Then $\vmax_{PP}(t,a)=maxTS(t,x)$. Let $\sigma \in \vmax_{PP}(t,a)$. 
\begin{itemize}
    \item $\sigma.wb^{t'',x}=\langle\,\rangle$ for all $t''\in \Tid$. For $(\sigma,\sigma')\in T_{PP}(t,\fence)$ we still have $\sigma'.wb^{t'',x}=\langle\,\rangle$ for all $t''\in \Tid$ and so $\sigma'\in \vmax_{PP}(t',a)$. Hence $\sigma\in \wlp(T_{PP}(t,\fence),\vmax_{PP}(t',a))$. 
    \item $\sigma.wb^{t,x}\neq \langle\,\rangle$. Let $(\sigma,\sigma')\in T_{PP}(t,\fence)$. For $\overset{\fence,t}{\leadsto}_{PP}$, $wb^{t,x}$ has to be empty. Since $\sigma \in maxTS(t,x)$ we need to flush $wb^{t'',x}$ for all $t''\in \Tid$ before we can flush $wb^{t,x}$. Then $\sigma'.wb^{t'',x}=\langle\,\rangle$ for all $t''\in \Tid$, $\sigma'\in \vmax_{PP}(t',a)$ and $\sigma\in \wlp(T_{PP}(t,\fence),\vmax_{PP}(t',a))$.
\end{itemize}

\end{document}